\documentclass{optica-article}

\journal{opticajournal} 

\articletype{Research Article}

\usepackage{lineno}
\usepackage{nicefrac}
\usepackage{braket}
\usepackage{subfig}

\def\figref#1{Fig.~\ref{#1}}

\def\emath{\mathrm{e}}

\newcommand{\abs}[1]{\left| #1 \right |}
\newcommand{\hc}{\mathrm{h.c.}}

\begin{document}

\title{Loss-aware optimization of realistic GKP grid states}

\author{Michal Matul\'ik,\authormark{1,*} Petr Marek,\authormark{1}}

\address{\authormark{1}Department of Optics, Palack\'y University, 17. listopadu 1192/12, 771 46 Olomouc, Czech Republic}

\email{\authormark{*}matulik@optics.upol.cz} 


\begin{abstract*}
Realistic preparation of Gottesman-Kitaev-Preskill (GKP) states, the prime candidates for quantum computation with bosonic systems, is inevitably burdened by photon loss. We show how taking this loss into account can lead to more robust quantum states, thereby diminishing its negative influence. Using GKP squeezing as an operational figure of merit, we formulate a loss-aware optimization procedure based on the ground state representation of GKP states.  Numerical simulations for square, hexagonal, and asymmetric GKP states show that the optimal preparation generally requires a different initial grid spacing than that of the desired ideal code state. For symmetric and hexagonal lattices, the optimal parameters asymptotically follow the simple scaling relation. The optimized states exhibit better GKP squeezing and greater resilience to losses. These results provide a simple and experimentally relevant strategy for improving realistic photonic GKP state preparation in the presence of known photon loss.

\end{abstract*}

\section{Introduction}

Nowadays, many physical platforms are considered for universal quantum computing~\cite{arute2019,zhong2020,yang2020,pogorelov2021}. Photonic technologies promise significant advantages over other approaches, such as room temperature computation and ease of scalability and networking~\cite{yokoyama2013,larsen2019,aghaeerad2025}. Nevertheless, like all platforms, it needs to deal with loss and noise. In order to mitigate the errors caused by these imperfections, quantum error correction is the most prominent strategy for fault-tolerant quantum computation~\cite{gottesman1997,menicucci2014}. The Bosonic codes it employs encode logical qubits into states from an infinite-dimensional Hilbert space~\cite{terhal2015,joshi2021}, the most common examples being the Gottesman-Kitaev-Preskill~(GKP) qubit~\cite{gottesman2001}, cat codes~\cite{bergmann2016}, or binomial codes~\cite{michael2016}. Among these, the GKP code holds a prominent position thanks to its ability to efficiently mitigate errors only with the help of feasible Gaussian operations \cite{baragiola2019}.

However, Gaussian operations and measurements alone are not sufficient for the universality of a quantum computation~\cite{lloyd1999,mari2012}. They are considered a 'free' resource in the resource theory of non-Gaussianity~\cite{walschaers2021}, even though, in practical scenarios, they have associated costs in the form of loss and added noise~\cite{miwa2014,asavanant2019,kashiwazaki2020,yamashima2025}. The main power of the Bosonic codes, therefore, lies in the non-Gaussian quantum states used to encode the information. For example, the GKP qubit is effective at mitigating errors due to its periodic, sharp-peaked phase space grid structure, which enables easy detection and correction of small loss and displacement errors~\cite{baragiola2019}. Experimental preparation of such states is one of the main bottlenecks for the realization of quantum computation~\cite{niset2009}. High quality GKP states have been achieved in trapped-ion~\cite{fluhmann2019,matsos2024} and superconducting circuit platforms~\cite{campagneibarcq2020,lachancequirion2024}, but these do not allow for easy networking. On the other hand, the traveling light platform with its high networking potential \cite{yokoyama2013,larsen2019,aghaeerad2025} currently needs to prepare the states by employing conditioning photon number measurements~\cite{fiurasek2005,vasconcelos2010,eaton2019,provaznik2020,grygar2022,kawasaki2022,endo2023,konno2024} within a boson sampling circuit \cite{tzitrin2020,hanamura2026}and the produced states are therefore limited both by the success rate and the overall loss ~\cite{konno2024,larsen2025}. It is also, in principle, possible to generate GKP states nearly deterministically, by exploiting the third order optical nonlinearity \cite{yanagimoto2023}, but the available coupling parameters are presently orders of magnitude lower than what is necessary.

No matter what the preparation circuit, the experimentally prepared GKP states will always be only approximations of the ideal nonphysical qubits. Consequently, characterizing the quality of such approximate states is an important problem both in the design of the circuits and their experimental evaluation. Fidelity is generally easy to evaluate, but it is not practical in this case because the ideal states are unphysical. The pair of effective squeezing parameters \cite{weigand2018} is better at describing the two independent quadratures, but it is more difficult to evaluate. The recently introduced GKP squeezing ~\cite{marek2024,hanamura2026,kuchar2026} takes the best from both approaches by providing Hermitian positive semi-definite operators whose ground states are GKP states in various topologies. Expectation values of these operators define their non-linear squeezing~\cite{marek2024,kala2022,brauer2025,brauer2026,matulik2026}, determine the quality of the states similarly to \cite{weigand2018}, and can be used to certify their non-Gaussianity or even stellar rank \cite{provaznik2026}. And, since it is easy to evaluate, it can be straightforwardly used to optimize the state preparation procedures  \cite{hanamura2026}.

In this paper, we focus on improving the quality of GKP states in realistic scenarios with inherent loss. For GKP states, the loss shrinks the grid they exist on and adds the noise floor \cite{ivan2011}. We show that known losses can be taken into account and compensated for by preparing GKP on a slightly different grid. We find the optimal grid parameters leading to the minimal value of the GKP squeezing~\cite{marek2024}, and find the respective states as eigenstates of the GKP operator. We do consider neither the dephasing noise, for GKP codes are not sufficiently robust against it~\cite{wang2026}, nor the additive thermal noise, which does not change the grid and therefore leaves the optimal GKP states unchanged. 

The paper is organized into two major sections, starting with a general overview of the GKP squeezing and its optimization in realistic scenarios in the \hyperref[sec:2]{second section}, followed by the presentation of numerical results. In the \hyperref[sec:dis]{third section}, we discuss the obtained results.

\section{Optimization protocol}\label{sec:2}
The ideal GKP qubit state $\ket{0_L}$ is defined as the infinite superposition of quadrature eigenstates,~\cite{gottesman2001,marek2024,tzitrin2020,bourassa2021}
\begin{equation}\label{eq:gkp_0L}
    \ket{0_L} \propto \sum_{s \in Z } \ket{x= 2s \sqrt{\pi}}.
\end{equation}
The state $\ket{0_L}$ is the simultaneous eigenstate of displacement operators $\hat{D}_p = \hat{D}\big(\imath \textstyle \sqrt{\frac{\pi}{2}}\big)$ and $\hat{D}_{x} = \hat{D}(\sqrt{2 \pi  })$ where
\begin{equation}
    \hat{D}(\alpha) = \emath^{\alpha \hat{a}^{\dagger} - \alpha^* \hat{a}}, \; \; \; \; \alpha = \textstyle \frac{1}{\sqrt{2}} (x_0 + \imath p_0), \; \; \; \; x_0,p_0 \in \mathbb{R}.
\end{equation}
Here $\hat{a}^{\dagger}$ and $\hat{a}$ denote the creation and annihilation operators, respectively \cite{leonhardt}. They satisfy the commutation relation $[\hat{a},\hat{a}^{\dagger}] = 1$. The state \eqref{eq:gkp_0L} is also the ground state of the positive semidefinite operator~\cite{marek2024}
\begin{align}
    \hat{O}\Big(\textstyle \frac{\sqrt\pi}{2}, \sqrt{\pi}\Big) &= 2 \sin^2 \big(\hat{x} \textstyle{ \frac{\sqrt\pi}{2}} \big) + 2 \sin^2 (\hat{p} \sqrt{\pi} ) = \frac{1}{2} \big( 4 - \hat{D}_x - \hat{D}_x^{\dagger} - \hat{D}_p - \hat{D}_p^{\dagger} \big).
\end{align}

Any other topology of GKP-like states is only a Gaussian operation away, and it can be found as the ground state $\hat{\rho}(\mathbf{g})$, $i \in \{1,2\}$ of the general positive semidefinite operator~\cite{marek2024}
\begin{align}\label{eq:THEoperator}
    \hat{O}(\mathbf{g}) &= 2\sin^2(g_1 \hat{x} - g'_{1} \hat{p}) +2 \sin^2 ( g_2 \hat{p} - g'_2 \hat{x}) \\ & = \frac{1}{2} \bigg( 2 \hat{I} - \hat{D}\big(\sqrt{2}(g_1 + \imath g'_1)   \big)  - \hat{D}\big(\sqrt{2}(g'_2 + \imath g_2) \big) + \hc  \bigg),
\end{align}
where 
\begin{align}\label{eq:kmatrix}
\mathbf{g} = \left(
\begin{array}{cc}
  g_1 & g_1' \\
  g_2 & g_2' 
\end{array}\right)
\end{align}
is a renormalized symplectic matrix with $|\mathbf{g}| = \frac{\pi}{2}$ defining the spacing of the initial grid as well as the transformation corresponding to the alteration of the topology. Note that this renormalization is necessary for achieving perfect local squeezing in both quadratures. Operator~\eqref{eq:THEoperator} can also be used as a way to evaluate the quality of imperfect GKP states. For an arbitrary state $\hat{\rho}$, we can define the \textit{GKP squeezing} $\xi$ as the expectation value of  operator \eqref{eq:THEoperator},~\cite{marek2024},
\begin{align}\label{eq:ksi}
    \xi \equiv \text{Tr} [\hat{\rho} \hat{O}(\mathbf{g}) ].
\end{align}
The expectation value is equal to zero only when the expectation value of the displacement operators is 1. Thus, the GKP squeezing $\xi$ is 0 only for translationally invariant states, which are the ideal GKP states with topology given by $\mathbf{k}$. On the set of Gaussian states, the minimum achievable of $\xi$ is~1~\cite{marek2024}. Detecting smaller values, therefore, verifies the non-Gaussian nature of the state, and the GKP squeezing can be considered as a witness of non-Gaussianity and even of stellar rank~\cite{kala2022,brauer2025,brauer2026,matulik2026,provaznik2026}.

Ideal GKP states are nonphysical, their construction would require infinite energy~\cite{gottesman2001}. The operator~\eqref{eq:THEoperator} can be used to numerically find finite-energy approximations on finite-dimensional Fock spaces~\cite{marek2024} by finding the ground state of an operator obtained by projecting~\eqref{eq:THEoperator} to a finite dimensional subspace. This procedure leads to a finite number of peaks in the phase space description of the state, the number depending on the chosen Hilbert space dimension $N$ of the cut-off, and it is directly tied to stellar rank hierarchy~\cite{chabaud2020,provaznik2026}. The eigenvalues of the ground states are nonzero in finite-dimensional Hilbert spaces, the quality of such ground states is determined by the eigenvalue, which is the alternative definition of GKP squeezing $\xi$. Interestingly enough, this procedure can also be used to find states that do not exist on the ideal grid and do not satisfy $|\mathbf{g}| = \frac{\pi}{2}$. They can still be found as eigenstates of the operator $\hat{O}(\mathbf{g})$, but we will denote their grid parameters by $\mathbf{k}$ to avoid confusion.  In this scenario, the displacements in~\eqref{eq:THEoperator} do not commute, and $\xi$ can never attain the value of 0, even on an infinite dimensional Hilbert space, but the states still are and can be verified as non-Gaussian.

\begin{figure}[hbt]
    \includegraphics[width = \textwidth]{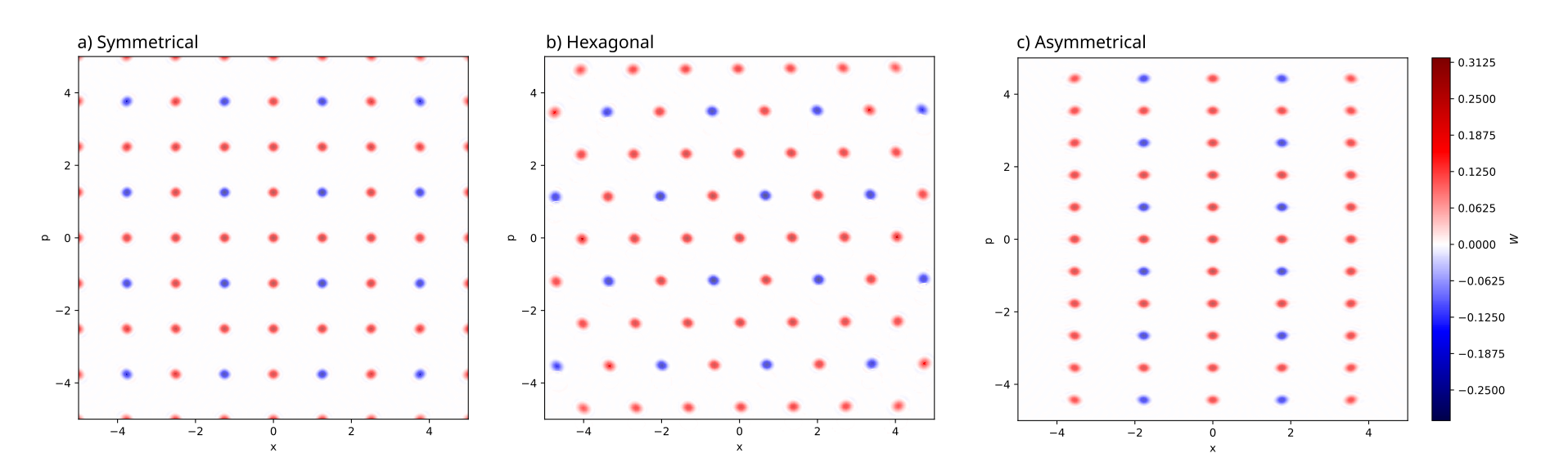}
    \caption{Wigner functions of a) square grid $\ket{0_S}$, b) hexagonal grid $\ket{0_H}$, and c) asymmetrical grid $\ket{0_L}$. Hexagonal grid is rotated by $\phi = -\sqrt{\pi/8 } (3^{1/4} + 3^{-1/4})$.}
    \label{fig:grids}
\end{figure}

One of the processes that can change the grid coefficients of the GKP states is photon loss, which is the main source of imperfections in the photonic platform. Loss can be modeled using an unbalanced beam splitter with intensity transmittance $0 < \eta^2 < 1$. The loss channel has an equivalent decomposition using the Kraus formalism~\cite{ivan2011}:
\begin{align}\label{eq:loss}
    \hat{\rho} (\eta) &= \sum_{l =0} ^{\infty} \hat{A}_{l} \hat{\rho} \hat{A}_{l}^\dagger, \; \; \; \; \hat{A}_{l} = \sqrt{\frac{(1-\eta^2)^l}{l!}} \eta^{\hat{a}^{\dagger} \hat{a}} \hat{a}^l.
\end{align}
The loss channel is trace-preserving, and the Kraus operators satisfy the completeness relation:
\begin{equation}\label{eq:completeness_kraus}
    \sum_l \hat{A}^\dagger_l \hat{A}_l = \hat{I}.
\end{equation}
The loss can be taken into account when designing state preparation circuits. Instead of analyzing an ideal lossless scenario, looking for a circuit designing a specific ideal quantum state or one that minimizes a specific cost function~\cite{hanamura2026}, one may look for a circuit that does that after the loss has taken its toll. In our case, where we seek GKP states characterized by their GKP squeezing, the optimization problem can be formally expressed as the search for
\begin{equation}\label{eq:minimization}
\begin{split}
    \xi_{\text{opt}} = \min_{\hat{\rho}}\text{Tr}[\hat{\rho}(\eta) \hat{O}(\mathbf{g})],
    \end{split}
\end{equation}
where $\mathbf{g}$ is a parameter with $|\mathbf{g}| = \frac{\pi}{2}$ representing the ideal grid. The minimization can be simplified by transferring the loss channel from the state to the operator. This can be done by using a coherent basis representation because the amplitude of coherent states under photon loss~\eqref{eq:loss} is reduced by $\eta$: $ \ket{\alpha} \rightarrow \ket{\eta \alpha}$,~\cite{ivan2011}.
Every state can be represented as:
\begin{equation}\label{eq:coherent_basis}
    \hat{\rho} = \int P(\alpha) \ket {\alpha} \bra{\alpha} d^2 \alpha,
\end{equation}
where $P(\alpha)$ is the P-function \cite{leonhardt}.  The identity operator in \eqref{eq:THEoperator} is invariant under loss due to the completeness relation \eqref{eq:completeness_kraus}. The displacement operators, however, transform non-trivially. We use the cycling property of the trace and coherent basis representation \eqref{eq:coherent_basis} to obtain 
\begin{equation}\label{eq:minimization2}
\begin{split}
    \xi_{\text{opt}} = \min_{\hat{\rho}}\text{Tr}[\hat{\rho} \hat{O}(\mathbf{g},\eta)],
    \end{split}
\end{equation}
where
\begin{equation}
\begin{split}\label{eq:Oeta}
       \hat{O}(\mathbf{g},\eta) = 2&- \textstyle\frac{1}{2}\emath^{-\abs{g_1 + \imath g'_1}^2(1-\eta^2)}\big(\hat{D}(\eta \sqrt2 (g_1 + \imath g'_1) ) + \hc\big)  \\ &- \textstyle\frac{1}{2}\emath^{-\abs{g'_2 + \imath g_2 }^2(1-\eta^2)}\big(\hat{D}(\eta \sqrt2(\imath  g_2+ g'_2) ) + \hc\big) .
       \end{split}
\end{equation}
The achievable minimum is then equal to the least eigenvalue of operator~\eqref{eq:Oeta}, and the quantum state needing to reach it can be found as the corresponding ground state. The operator \eqref{eq:Oeta} is from an infinitely dimensional Hilbert space. To get a practically obtainable quantum state, we will again need to project it to a finite dimension before obtaining the eigenstate, and the quality of the obtained state will depend on the chosen dimension. The cost, however, is that the minimal value can no longer be equal to 0. However, there is a caveat, and it is the degeneracy of the ground states of the operator \eqref{eq:Oeta}. There appears to be a degeneracy threshold that depends on the chosen finite dimension of the projection and on the values of $\mathbf{g}$ and $\eta$.  We will see that if the operator has a non-degenerate ground state, the ground state exhibits a similar GKP-like structure, but this time on a shrunk grid. Furthermore, to see the robustness of the solution, we will also evaluate the GKP squeezing for a broader family of GKP-like states, looking at
\begin{equation}\label{eq:minimization3}
\begin{split}
    \xi(\mathbf{k},\mathbf{g}) = \text{Tr}[\hat{\rho}(\mathbf{k}) \hat{O}(\mathbf{g},\eta)],
    \end{split}
\end{equation}
where $\hat{\rho}(\mathbf{k})$ is the least eigenvalue eigenstate of operator \eqref{eq:THEoperator} again cut down to the chosen dimension. In the following, we will perform the aforementioned analysis on three important topologies of GKP states, see \figref{fig:grids}, each with a different shape of the grid. All can be found as eigenstates of \eqref{eq:THEoperator} with proper selection of parameters $\mathbf{g}$. 

Square GKP state $\ket{0_S}$ \cite{walshe2020,hastrup2021} with
\begin{equation}\label{eq:params_sym}
    g \equiv g_{1,2} = \sqrt{\pi/2}, \; \; \; \; g'_{1,2} = 0.
\end{equation}  
illustrated in \figref{fig:grids}a, hexagonal $\ket{0_H}$ \cite{grimsmo2020,hastrup2021} with
\begin{align}\label{eq:params_hex}
     g_+ &\equiv g_{1}=g_{2} = \sqrt{\pi/8 } (3^{1/4} + 3^{-1/4}), \\ g_- &\equiv g'_{1}=g'_2 = \sqrt{\pi/8 } (3^{1/4} - 3^{-1/4}).
\end{align}
seen in \figref{fig:grids}b, hexagonal grid is rotated by $\phi = -\sqrt{\pi/8 } (3^{1/4} + 3^{-1/4})$ with rotation operator defined as $\hat{R} = \exp{(\imath \phi \hat{n})}$.
The logical GKP qubit $\ket{0_L}$ with the asymmetric rectangular grid given by
 \begin{equation}\label{eq:params_asym}
    g_{1} = \sqrt{\pi}/2, \; \; \;  g_{2} = \sqrt{\pi}, \; \; \; g'_{1,2} = 0.
\end{equation}
 seen in \figref{fig:grids}c. 

We numerically evaluate the minimal GKP squeezing \eqref{eq:minimization} or \eqref{eq:minimization2} for fixed parameters $\mathbf{g}$, which are given above, in dependence on the losses $\eta^2$ and the dimension of the available Hilbert space. We compare the optimized $\xi_{\mathrm{opt}}$ with the unoptimized $\xi$, which was obtained as from \eqref{eq:ksi} with $\hat{\rho} \equiv \hat{\rho}(\eta).$ We also evaluate GKP squeezing \eqref{eq:minimization3} for various grid parameters $\mathbf{k}$. In all cases, we also compute the difference between the first two minimal eigenvalues $\Delta$ of the \eqref{eq:Oeta}, 
\begin{equation}\label{eq:Delta}
\Delta = \lambda_1 - \lambda_0.
\end{equation}
Due to the phenomenon of avoided crossing, the difference $\Delta$ is not going to reach zero \cite{cohen}. The avoided crossing happens when two eigenvalues of two eigenstates get closer to each other but never intersect. In that critical moment, the two eigenstates are mixed with each other \cite{cohen}. 
We define the avoided-crossing bound as $\Delta < 10^{-3}$ to account for numerical tolerance. Under this bound, the states do not possess a GKP-like structure on the defined geometry of the grid. The minimal eigenstates still possess periodic structure in phase space, they can be a mixture of the ground state and the first excited state. 

\begin{figure}[hbt]
    \includegraphics[width = \textwidth]{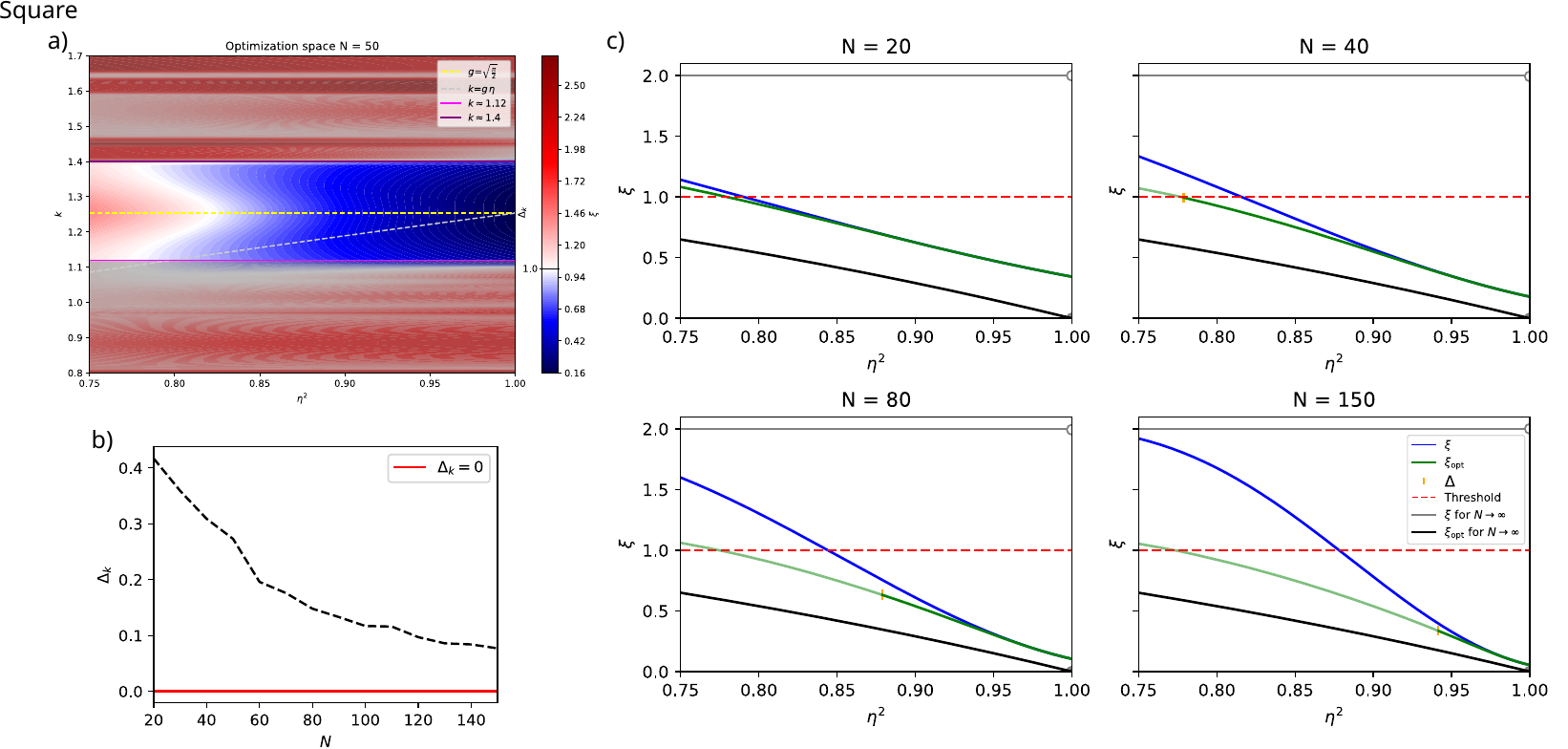}
    \caption{Optimization for square grid. Panel a) shows the typical optimization space for $k$ for various losses $\eta^2$ for $N = 50$, see \eqref{eq:minimization3}. The yellow dashed line indicates the value $g = \sqrt{\pi/2}$. The gray dashed line indicates the value $k = \textstyle  \sqrt{\frac{\pi}{2}}\eta$. The purple curves denote the degeneracy bound; the ground states outside the region $\Delta_k$ are degenerate ground states of the operator \eqref{eq:Oeta}. Panel b) shows the size of the interval $\Delta_k$ as a function of the dimension $N$ of the operator \eqref{eq:THEoperator} for various $k$.
    Panel c) shows $\xi$ for optimized and non-optimized grid parameters across four examples for the chosen $N = \{20,40,80,150\}$. The blue line corresponds to GKP squeezing $\xi$, which was computed for a lossy symmetric GKP state obtained as the ground state of the operator with fixed grid parameters $g = \sqrt{\pi/2}$ propagated through the loss channel, and the green curve shows minimal eigenvalues $\xi_{\text{opt}}$ of the operator \eqref{eq:Oeta} for given $\eta$ and $g$. The green curve is divided by the avoided crossing threshold $\Delta$, denoted by the orange mark. The eigenstates in dark green have the same geometry of the grid, but the grid itself is shrunken. Ground states denoted in light green have a different grid geometry. The red dashed line indicates the Gaussian threshold. The black line represents $\xi_{\mathrm{opt}}$ in an infinite-dimensional Hilbert space. The gray line represents a discontinuous function $\xi$ in the infinite-dimensional case.} 
    \label{fig:sym}
\end{figure}
\subsection{Square grid}
For the square grid, \figref{fig:grids}a with parameters \eqref{eq:params_sym}, the losses modify both quadratures in the same way. The optimal state can therefore be found as the minimal eigenvalue eigenstate of $\hat{O}(\mathbf{k})$ with $k_{1,2} = k = g \eta$, $g = \textstyle \sqrt{\frac{\pi}{2}}$. 
Nevertheless, in \figref{fig:sym}a we show the dependence of the GKP squeezing for different values of $k$, different values of $\eta$, and the Hilbert space cut-off dimension $N = 50$, see \eqref{eq:minimization3}. We can see that the minimum does really follow the analytical optimum, see gray dashed line, but also that the optimum is fairly robust, achieving similar values for $1.1 < k < 1.4$. We can also see that for losses with $\eta^2 \sim 0.85$ it might also be beneficial to prepare states on a larger grid, with $k \approx g/\eta$, as this provides almost the same value as the optimal solution. We can also see that the optimum crosses the avoided crossing bound at $\eta^2 \sim 0.8$, the bound is indicated by the purple line. The horizontal lines were computed from the degeneracy of ground states of $\hat{O}(k)$ and are not dependent on $\eta$. The crossing means that the optimal states minimizing GKP squeezing $\xi_{\mathrm{opt}}$ for lower values of $\eta^2$ do not have GKP-like structure because of the avoided crossing phenomenon. In this case, we also define the difference between upper and lower avoided crossing bounds $\Delta_k$. We plot the size of the interval $\Delta_k$ as a function of dimension $N $ in \figref{fig:sym}b. The size of the interval shrinks as the dimension increases.
The optimal GKP squeezing relative to $\eta^2$ is then shown in \figref{fig:sym}c for $N = \{20,40,80,150\}$. In each figure, the blue line corresponds to GKP squeezing, $\xi$, which was computed for a lossy GKP state obtained as the ground state of the operator with fixed grid parameters $g = \sqrt{\pi/2}$ propagated through the loss channel, and the green curve shows minimal eigenvalues $\xi_{\text{opt}}$ of the operator \eqref{eq:Oeta} for given $\eta$~and~$g$.  We can see that with the increasing dimension $N$ the ideal states marked by blue lines exhibit better GKP squeezing for $\eta=1$, but also more rapid deterioration due to loss. This is even more visible in the limit $N\rightarrow \infty$, which is a discontinuous function 
\begin{equation}\label{eq:dependece_N_infity_nonoptimal}
    \xi_{\text{initial}}=
\begin{cases}
2, & 0 \le \eta^2 < 1,\\
0, & \eta^2 = 1,
\end{cases}
\; \; \;  \text{for }N \rightarrow \infty.
\end{equation}
indicated by the gray line. On the other hand, the deterioration of loss optimized states with the optimal $k$, marked by the green lines, deteriorates more slowly, especially for larger dimensions. The effect is again best observed in the limit of $N \rightarrow \infty$, which can be analytically found to be
\begin{equation}\label{eq:dependence_N_infty}
    \xi_{\text{opt}} = (2 - 2 \emath^{-\abs{g}^2 (1-\eta^2)}), \; \; \; \text{for }N \rightarrow \infty,
\end{equation}
and is marked by the black lines. The avoided crossing bound $\Delta$ is denoted by orange vertical marks. The ground states defined on the shrunken grid are denoted in dark green, and the different grid geometry states are denoted in light green. Therefore, only the states corresponding to dark green GKP squeezing have a similar structure to non-optimized states but on a shrunken grid. The ground states under avoided crossing threshold still possess periodic structure in phase space, but they do not remind us of any known non-Gaussian state.

 \begin{figure}[hbt]
    \includegraphics[width = \textwidth]{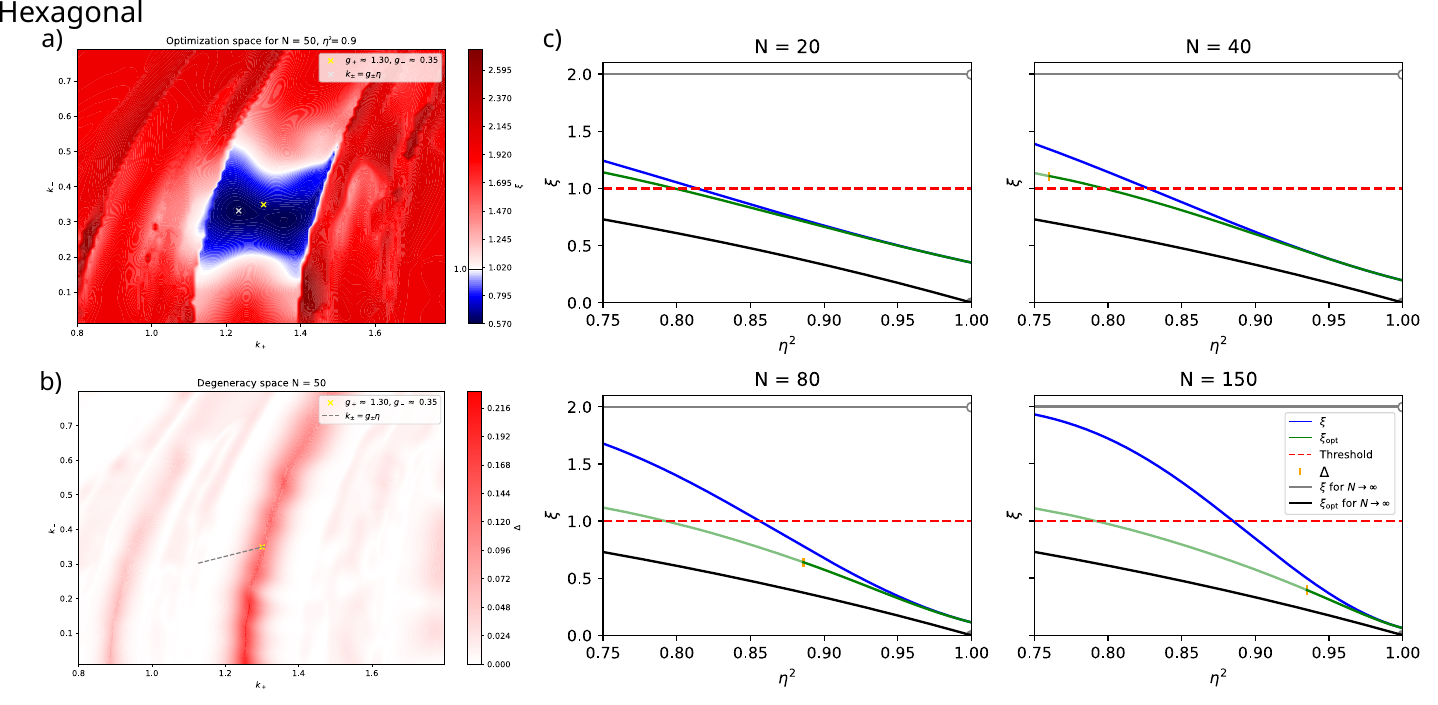}
    \caption{Hexagonal geometry of the grid. Panel a) shows the typical optimization space for $k_+$ and $k_-$ for fixed $\eta^2 =0.9$ for $N = 50$. The yellow mark indicates the value $g_+  = \sqrt{\pi/8 } (3^{1/4} + 3^{-1/4})$ and $g_-  = \sqrt{\pi/8 } (3^{1/4} - 3^{-1/4})$. The gray mark indicates the value $k_{\pm} = g_{\pm}\eta$. Panel~b) shows difference $\Delta$ between the two minimal eigenvalues of the operator \eqref{eq:THEoperator}. The yellow mark once again indicates the value $g_+  = \sqrt{\pi/8 } (3^{1/4} + 3^{-1/4})$ and $g_-  = \sqrt{\pi/8 } (3^{1/4} - 3^{-1/4})$. The gray dashed line indicates the function $k_{\pm} = g_{\pm}\eta$ for $\eta^2 \in [0.75,1].$ Panel c) shows $\xi$ for optimized and non-optimized grid parameters across four examples for the chosen $N = \{20,40,80,150\}$. The blue line corresponds to GKP squeezing $\xi$, which was computed for a lossy symmetric GKP state obtained as the ground state of the operator with fixed grid parameters $g_{\pm}$ propagated through the loss channel, and the green curve shows minimal eigenvalues $\xi_{\text{opt}}$ of the operator \eqref{eq:Oeta} for given $\eta$ and $g_{\pm}$. The green curve is divided by the avoided crossing threshold $\Delta$, denoted by the orange mark. The eigenstates in dark green have the same grid geometry, but the grid itself is shrunk. Ground states denoted in light green have a different grid geometry. The red dashed line indicates the Gaussian threshold. The black line represents $\xi_{\mathrm{opt}}$ in an infinite-dimensional Hilbert space. The gray line represents a discontinuous function $\xi$ in the infinite-dimensional case.}
    \label{fig:hex}
\end{figure}
\subsection{Hexagonal grid}
The hexagonal grid, \figref{fig:grids}b with parameters \eqref{eq:params_hex}, is also symmetric, meaning that the optimal state can again be found as the minimal eigenvalue eigenstate of $\hat{O}(\mathbf{k})$ with $\mathbf{k} = \mathbf{g} \eta$ and \eqref{eq:params_hex}. Nevertheless, to see the parameter dependence, we also perform a general optimization where we consider the parameters $k_+$ and $k_-$ to be free. \figref{fig:hex}a shows the 2D optimization space for fixed losses $\eta^2 = 0.9$ and for $N = 50$. The non-optimized parameters $g_+  = \sqrt{\pi/8 } (3^{1/4} + 3^{-1/4})$ and $ g_-  = \sqrt{\pi/8 } (3^{1/4} - 3^{-1/4}) $ are marked by yellow cross, while the optimum $k_{\pm} =  g_{\pm} \eta$ is marked by a gray one. We can see that the range of parameters is better than the initial one, and is again quite broad.  Please note that there is a second, equivalent blue region due to the invariance under the exchange of $g_+ \leftrightarrow g_-$. \figref{fig:hex}b shows difference $\Delta$ between the two minimal eigenvalues of the operator $\hat{O}(\mathbf{k})$. The yellow mark once again indicates the value $g_+  = \sqrt{\pi/8 } (3^{1/4} + 3^{-1/4})$ and $g_-  = \sqrt{\pi/8 } (3^{1/4} - 3^{-1/4})$. The gray dashed line indicates the function $k_{\pm} = g_{\pm}\eta$ for $\eta^2 \in [0.75,1].$ The avoided-crossing boundary $\Delta \sim 0$ is represented by white regions. The gray dashed line indicates how the optimal state is moving with respect to known losses, and once again, the optimum enters the white region; therefore, for some non-zero losses, the ground states of the operator \eqref{eq:Oeta} are degenerate. \figref{fig:hex}c then shows the GKP squeezing for optimized and non-optimized grid parameters across four examples for the chosen $N = \{20,40,80,150\}$. In each figure, the blue line corresponds to GKP squeezing, $\xi$, which was computed for a lossy GKP state obtained as the ground state of the operator with fixed grid parameters $g_{\pm}$ propagated through the loss channel,  and the green curve shows minimal eigenvalues $\xi_{\text{opt}}$ of the operator \eqref{eq:Oeta} for given $\eta$ and $g_{\pm}$.  We can see that with the increasing dimension $N$ the ideal states marked by blue lines exhibit better GKP squeezing for $\eta^2=1$, but also more rapid deterioration due to loss. The gray and black lines then again show the limit of $N\rightarrow \infty$ for the non-optimized and optimized GKP state, respectively. They can be found analytically, for the initial non-optimized state, the GKP squeezing after decoherence is again given by \eqref{eq:dependece_N_infity_nonoptimal}, while for the optimized state it is
\begin{equation}\label{eq:dependence_N_infty_hex}
    \xi_{\text{opt}} = (2 - 2 \emath^{-(g_+^2 + g_-^2) (1-\eta^2)}), \; \; \; N \rightarrow \infty,
\end{equation}
Similarly to \figref{fig:sym}, we can see that optimizing the state significantly enhances its robustness against losses, especially for larger dimensions of the Hilbert space.

\begin{figure}[hbt]
    \includegraphics[width = \textwidth]{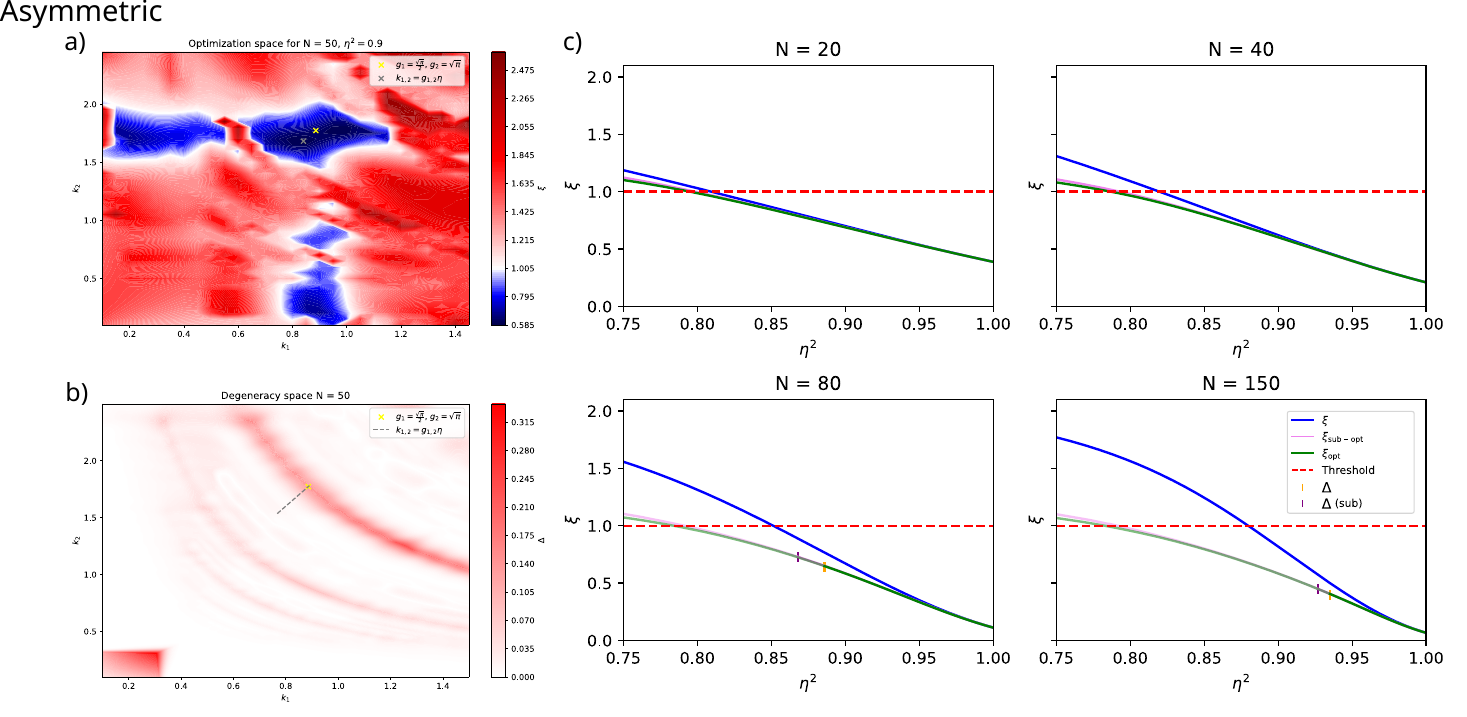}
    \caption {Asymmetric geometry of the grid. Panel~a) shows the typical optimization space for $k_1$ and $k_2$ for fixed $\eta^2 =0.9$ for $N = 50$. The yellow mark indicates the value $g_1  = \sqrt{\pi}/2$ and $g_2  = \sqrt{\pi } $. The gray mark indicates the value $k_{1,2} = g_{1,2}\eta$. Panel~b) shows difference $\Delta$ between the two minimal eigenvalues of the operator \eqref{eq:Oeta_mod}. The yellow mark once again indicates the values $g_1$ and $g_2$. The gray dashed line indicates the function $k_{1,2} = g_{1,2}\eta$ for $\eta^2 \in [0.75,1].$ Panel c) shows $\xi$ for optimized and non-optimized grid parameters across four examples for the chosen $N = \{20,40,80,150\}$. The blue line corresponds to GKP squeezing $\xi$, which was computed for a lossy symmetric GKP state obtained as the ground state of the operator with fixed grid parameters $g_{1,2}$ propagated through the loss channel, and the green curve shows minimal eigenvalues $\xi_{\text{opt}}$ of the operator \eqref{eq:Oeta} for given $\eta$ and $g_{1,2}$. The pink line indicates GKP squeezing of the operator \eqref{eq:Oeta}, but for the minimal eigenvalue eigenstates of $\hat{O}(\mathbf{k)}$, $\mathbf{k} = \mathbf{g} \eta$. The green curve is divided by the avoided crossing threshold $\Delta$, denoted by the orange mark. The eigenstates in dark green have the same geometry of the grid, but the grid itself is shrunken. Ground states denoted in light green have a different grid geometry. The same applies to the pink link, it is also divided by the purple vertical marker. The red dashed line indicates the Gaussian threshold.}
    \label{fig:nonsym}
\end{figure}
\subsection{Asymmetric grid}
Unlike the two previous cases, the logical grid \figref{fig:grids}c  with parameters \eqref{eq:params_asym} is not symmetric. The optimal state can still be found as the minimal eigenvalue eigenstate of \eqref{eq:Oeta}, but it does not exactly align with eigenstates of $\hat{O}(\mathbf{k})$ no matter the choice of $\mathbf{k}$. In other words, we do not know an exact analytical prescription for $\mathbf{k}$ as in the cases above. However, the approximate solution can be found. The operator \eqref{eq:Oeta} can be reformulated
\begin{equation}
    \begin{split}\label{eq:Oeta_mod}
       \hat{O}(\mathbf{g},\eta) = 2- \big(\hat{D}(\eta \sqrt2 (g_1 + \imath g'_1) ) + \hc\big)  - \big(\hat{D}(\eta \sqrt2(\imath  g_2+ g'_2) ) + \hc\big) + \delta \hat{O},
       \end{split}
\end{equation}
where $\delta \hat{O}$:
\begin{equation}
     \delta\hat{O}(\mathbf{g},\eta) = \textstyle\frac{c_1}{2}\big(\hat{D}(\eta \sqrt2 (g_1 + \imath g'_1) ) + \hc\big)  +\frac{c_2}{2}\big(\hat{D}(\eta \sqrt2(\imath  g_2+ g'_2) ) + \hc\big),
\end{equation}
with 
\begin{align}
    c_1 &= 1 - \emath^{-\abs{g_1 + \imath g'_1}^2(1-\eta^2)}, \; \; \; \; c_2  = 1 - \emath^{-\abs{g'_2 + \imath g_2 }^2(1-\eta^2)}.
\end{align}
When $\eta^2 \sim 1$ we can neglect $\delta \hat{O}$. The optimal state for minimizing GKP squeezing of \eqref{eq:Oeta_mod} can be again found as the minimal eigenvalue eigenstate of $\hat{O}(\mathbf{k})$ with $\mathbf{k} = \mathbf{g} \eta$ and \eqref{eq:params_asym}.

The logical \figref{fig:nonsym} summarizes numerical results for an asymmetric grid \eqref{eq:params_asym}. To see the parameter dependence, we perform a general optimization where we consider the parameters $k_1$ and $k_2$ to be free. \figref{fig:nonsym}a shows the 2D optimization space for $k_1$ and $k_2$ for fixed losses $\eta^2 = 0.9$ and for $N = 50$. The optimization space is not smooth. The GKP squeezing of the specific state with non-optimized parameters $g_1  = \sqrt{\pi}/2$ and $ g_2 = \sqrt{\pi}$ is marked by a yellow cross, while the optimum $k_{1,2} = g_{1,2}\eta$ is marked by the gray one. \figref{fig:nonsym}b) shows the difference $\Delta$ between the two minimal eigenvalues of the operator $\hat{O}(\mathbf{k})$. The yellow mark once again indicated the values $k_1$ and $k_2$ specified above. The gray dashed line indicates the function $k_{1,2} = g_{1,2}\eta$ for $\eta^2 \in [0.75,1]$. The ground states of the operator \eqref{eq:Oeta_mod} are degenerate when the dashed line enters the white region. \figref{fig:nonsym}c shows GKP squeezing for optimized and non-optimized grid parameters across four examples for the chosen $N = \{20,40,80,150\}$. In each figure, the blue line corresponds to GKP squeezing, $\xi$, which was computed for a lossy GKP state obtained as the ground state of the operator with fixed grid parameters $g_{1,2}$ propagated through the loss channel,  and the green curve shows minimal eigenvalues $\xi_{\text{opt}}$ of the operator \eqref{eq:Oeta} for given~$\eta$ and~$g_{1,2}$. The pink line indicates GKP squeezing of the operator \eqref{eq:Oeta}, but for the minimal eigenvalue eigenstates of $\hat{O}(\mathbf{k)}$, $\mathbf{k} = \mathbf{g} \eta$. The pink and the green lines are essentially overlapping for $\eta^2>0.9$. 

As in previous examples, we can see that with the increasing dimension $N$ the ideal states marked by blue lines exhibit better GKP squeezing for $\eta=1$, but also more rapid deterioration due to loss. Similarly to \figref{fig:sym}, we can see that optimizing the state significantly enhances its robustness against losses, especially for larger dimensions of the Hilbert space. The avoided crossing bound of the two lowest eigenstates of the operator \eqref{eq:Oeta} is marked by the orange vertical line. The avoided crossing bound of the two lowest eigenstates of \eqref{eq:Oeta_mod} is denoted by the purple vertical line. Those two bounds are close to each other. We do not plot maximum GKP squeezing as in the previous cases because it is not obvious what the optimum eigenstates of the operator \eqref{eq:Oeta} are.
\begin{figure}[hbtp]
    \includegraphics[width = \textwidth]{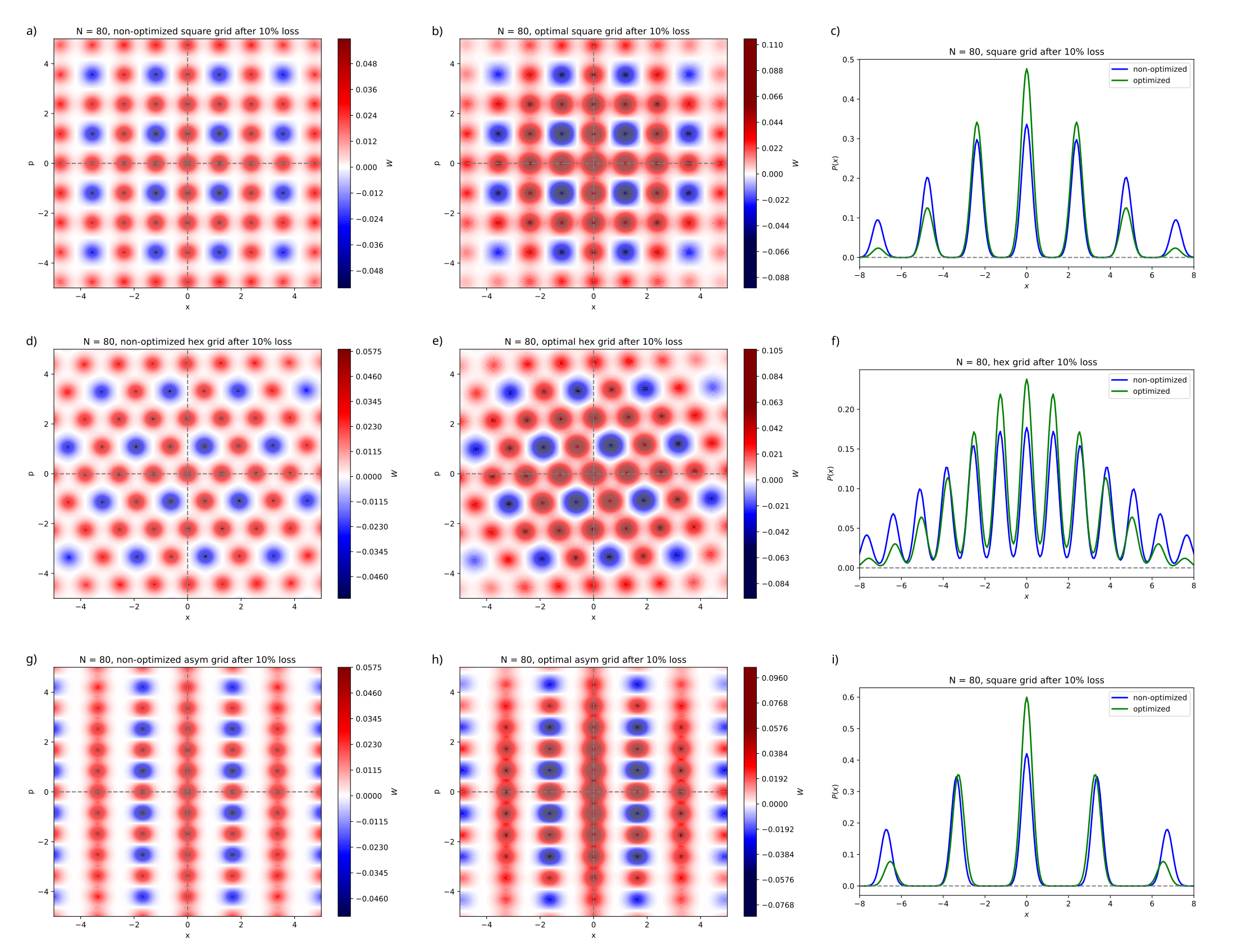}
    \caption{Figure shows nine panels, left column represents various grids obtained as eigenstates of the operator $\hat{O}(\mathbf{g)}$ propagated through the loss channel. The middle column represents states obtained as eigenstates of the operator $\hat{O}(\mathbf{k})$, where $\mathbf{k} = \eta \mathbf{g}$, and propagated through the loss channel. The panels a) and b) show a Wigner function of a square grid with $\mathbf{g}$ given by \eqref{eq:params_sym} on the Hilbert space spanned by Fock dimension $N = 80$. The panels d) and e) show the rotated hexagonal GKP state with $\mathbf{g}$ defined in \eqref{eq:params_hex} and cut at $N =80$. The panels g) and h) show an asymmetric grid with $\mathbf{g}$ given in \eqref{eq:params_asym}, the Hilbert space is limited to $N = 80$. The panels c), f), i) show a marginal distribution of the Wigner functions $P(x)$.}
    \label{fig:grid_transformation}
\end{figure}
\section{Discussion}\label{sec:dis}
\begin{figure}[hbtp]
    \includegraphics[width = \textwidth]{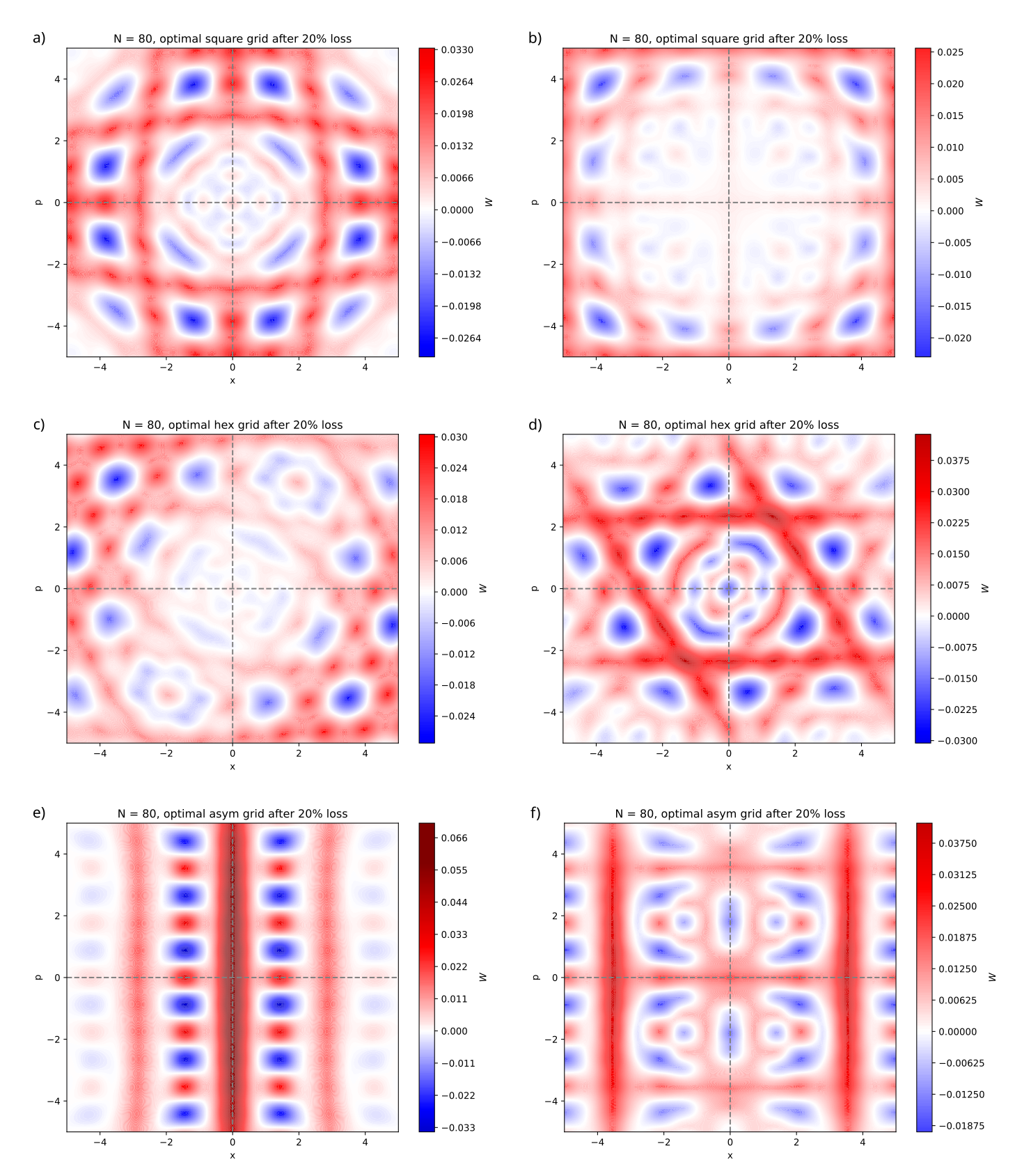}
    \caption{Figure of states under the avoided crossing boundary. Both columns represent states obtained as eigenstates of the operator $\hat{O}(\mathbf{k})$, where $\mathbf{k} = \eta \mathbf{g}$, and propagated through the loss channel. Each row represents two eigenstates corresponding to the two lowest energy levels, the numerical difference of eigenvalues $\Delta$ can be found in Tab. \ref{tab:delta_comparison}. The panels a) and b) show a Wigner function of a square grid with $\mathbf{g}$ given by \eqref{eq:params_sym} on the Hilbert space spanned by Fock dimension $N = 80$. The panels c) and d) show the rotated hexagonal GKP state with $\mathbf{g}$ defined in \eqref{eq:params_hex} and cut at $N =80$. The panels e) and f) show an asymmetric grid with $\mathbf{g}$ given in \eqref{eq:params_asym}, the Hilbert space is limited to $N = 80$. For all cases $\eta^2 = 0.9$.}
    \label{fig:grid_transformation_degen}
\end{figure}
We present Wigner functions of the non-optimized and optimized quantum states. 
\figref{fig:grid_transformation} shows nine panels. The figures in the left column represent various grids obtained as eigenstates of the operator $\hat{O}(\mathbf{g)}$ propagated through the loss channel. The middle column represents states obtained as eigenstates of the operator $\hat{O}(\mathbf{k})$, where $\mathbf{k} =  \mathbf{g}\eta $, and propagated through the loss channel. The chosen parameters $N=80$ and $\eta^2 = 0.9$ make sure that the eigenstates are above the avoided crossing threshold. This means that the ground states are not superpositions of excited states, therefore, the geometry of the grid is preserved. The first row of panels a) and b) shows a square grid with $\mathbf{g}$ given by \eqref{eq:params_sym} on a Hilbert space spanned by Fock dimension $N = 80$. The loss channel is represented by parameter $\eta^2 = 0.9$.
The panels d) and e) show the rotated hexagonal GKP state with $\mathbf{g}$ defined in \eqref{eq:params_hex}, and the Hilbert space is cut at $N =80$. Both states were propagated through the loss channel with intensity transmission $\eta^2 = 0.9$.  The panels g) and h) show an asymmetric grid with $\mathbf{g}$ given in \eqref{eq:params_asym}, the Hilbert space is limited to $N = 80$. The intensity transmission is $\eta^2 = 0.9$. The rightmost column represents a marginal probability distribution function, $P(x)$, and compares $P(x)$ of optimized and non-optimized grids in the left and middle columns. The optimized states for every GKP topology exhibit higher overall Wigner function maxima and lower minima. We can also see that the peaks farther from the origin are both smaller and shifted, meaning they suffer less from decoherence while maintaining the stabilizing properties. It is mostly seen in the marginal distribution of an asymmetric grid, see \figref{fig:grid_transformation}i. The minimization of the GKP squeezing leads to concentration of the Wigner function around the origin of the phase space. 
Table \ref{tab:gkp_sq_comparison} compares GKP squeezing $\xi$ and $\xi_{\mathrm{opt}}$ of six states a) - h) from \figref{fig:grid_transformation}. This clearly demonstrates the advantage of optimized states. 
\begin{table}[htbp]
\caption{GKP squeezing of optimized and non-optimized states from \figref{fig:grid_transformation}}
  \label{tab:gkp_sq_comparison}
  \centering
\begin{tabular}{ccc}
\hline
Grid  & $\xi$ & $\xi_{\text{opt}}$ \\
\hline
Square & 0.61 & 0.54 \\
Hex &  0.67 & 0.57 \\
Asym &  0.67 & 0.59 \\
\hline
\end{tabular}
\end{table}

\figref{fig:grid_transformation_degen} shows quantum states below avoided crossing threshold. Both columns represent two lowest eigenstates of the operator $\hat{O}(\mathbf{k})$, where $\mathbf{k} = \eta \mathbf{g}$, and propagated through the loss channel. The left column corresponds to ground state the and the right column to first excited eigenstate.  Each row represents two eigenstates corresponding to two lowest energy levels of different grid topology. The difference $\Delta$ between the energy levels of the eigenstates can be found in Tab. \ref{tab:delta_comparison}. \figref{fig:grid_transformation_degen}a and \figref{fig:grid_transformation_degen}b show a Wigner function of a square grid with $\mathbf{g}$ given by \eqref{eq:params_sym} on the Hilbert space spanned by Fock dimension $N = 80$. \figref{fig:grid_transformation_degen}c and \figref{fig:grid_transformation_degen}d show the rotated hexagonal GKP state with $\mathbf{g}$ defined in \eqref{eq:params_hex} and cut at $N =80$. The panels e) and f) show an asymmetric grid with $\mathbf{g}$ given in \eqref{eq:params_asym}, the Hilbert space is limited to $N = 80$. For all cases $\eta^2 = 0.8$.
\begin{table}[htbp]
\caption{Numerical difference $\Delta$ between eigenenergies of two lowest eigenstates defined from \figref{fig:grid_transformation_degen}}
  \label{tab:delta_comparison}
  \centering
\begin{tabular}{cc}
\hline
Grid  & $\Delta$ \\
\hline
Square & 6.147$\times 10 ^{-5}$ \\
Hex &  2.5$\times 10 ^{-3}$ \\
Asym &  1.6$\times 10 ^{-3}$ \\
\hline
\end{tabular}
\end{table}

\begin{figure}[hbtp]
    \includegraphics[width = \textwidth]{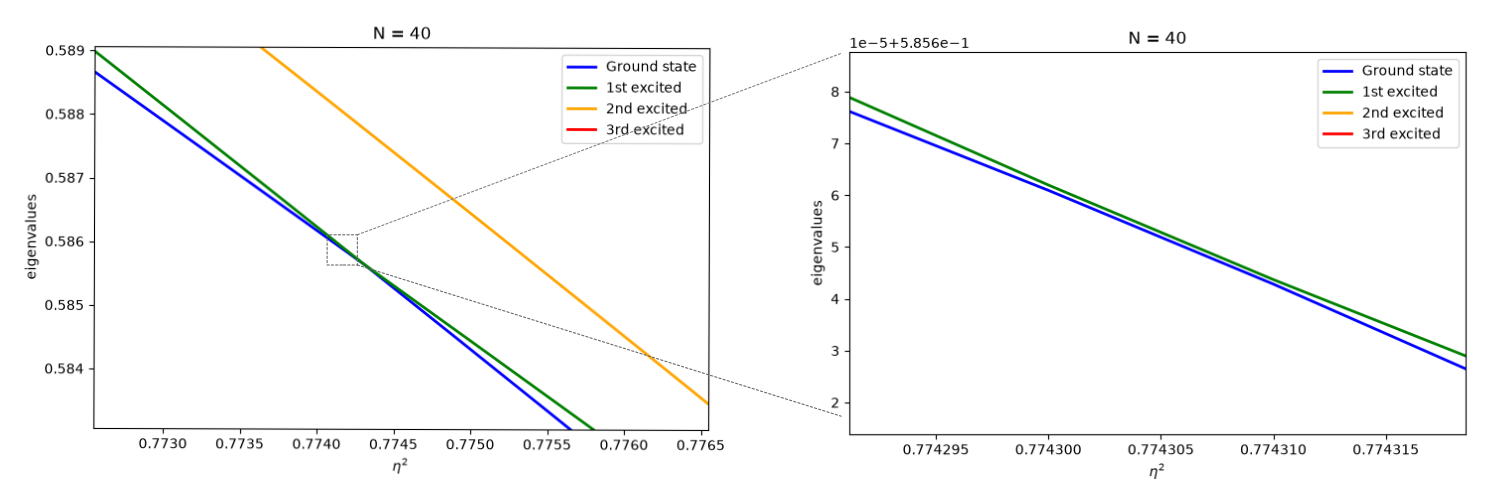}
    \caption{Figure shows eigenvalues of $\hat{O}(\mathbf{k})$ of the square grid for fixed $N = 40$ as a function of $\eta^2$. The ground energy is depicted by blue line, the first excited eigenvalue is in green. The third in orange. The right figure shows detail of the left one.}
    \label{fig:avoided_crossing}
\end{figure}

We explicitly plot first three eigenvalues of $\hat{O}(\mathbf{k})$ of the square grid as a function of $\eta^2$, see \figref{fig:avoided_crossing}. The dimension is fixed at $N = 40$. The ground energy and energy of the first excited state undergoes avoided crossing in the region of $\eta^2 \sim 0.7743$. This is in agreement with \figref{fig:sym}c. 

\section{Conclusion}
We have shown that GKP squeezing can be used to optimize the preparation of GKP states to account for known losses and that this approach does improve the resilience of the produced states. We have analyzed three significant types of GKP states, GKP states on a square, hexagonal, and rectangular grid, and shown that the optimal strategy is to prepare GKP-like states on a diminished grid. We have also shown that the GKP squeezing operator can be straightforwardly used to find these states, either as exact states in Fock representation that can be targeted by state preparation protocols, or by optimizing the state preparation protocols themselves with the GKP squeezing operator serving as a practical cost function \cite{hanamura2026}. Interestingly, for higher losses the phenomenon of avoided crossing plays a crucial role in the geometry of the eigenstates of the GKP squeezing operator.

While the effect is not significant for small dimensions of the Hilbert space and therefore not immediately relevant for near term experimental realizations with traveling light such as \cite{konno2024,larsen2025}, it becomes more pronounced as the dimension and energy of the states increases, meaning that the benefits could be seen in experiments involving other platforms, such as trapped ions \cite{fluhmann2019,matsos2024}, superconducting circuits \cite{campagneibarcq2020,lachancequirion2024}, or deterministic high order optical nonlinearities \cite{yanagimoto2023}.


\begin{backmatter}
\bmsection{Funding}
PM, MM acknowledge the financial support of the Czech Science Foundation (project 25-17472S), and European Unionâ's HORIZON Research and Innovation Actions under Grant Agreement no. 101080173 (CLUSTEC). MM acknowledges IGA-PrF-2026-005. MM acknowledges the Quantera project CLUSSTAR (8C24003) of MEYS, Czech Republic. Project CLUSSTAR has received funding from the European Union's Horizon 2020 Research and Innovation Programme under Grant Agreements No. 731473 and No. 101017733 (QuantERA). PM acknowledges a grant from the Programme Johannes Amos Comenius under the Ministry of Education, Youth and Sports of the Czech Republic reg. no. CZ.02.01.01/00/22\_008/0004649.

\bmsection{Acknowledgment}
MM acknowledges fruitful discussions with Jan Provazn{\'\i}k. The authors acknowledge the use of the Department of Optics' computational cluster and the use of Python libraries \cite{harris2020,virtanen2020}. The authors used OpenAI's ChatGPT-5.6 and Grammarly to obtain language-editing suggestions, including possible title formulation. The authors reviewed and take full responsibility for all content of the manuscript.
\bmsection{Disclosures}
The authors declare no conflicts of interest.

\bmsection{Data availability} The data underlying the results presented in this paper will be available in a public repository.
\end{backmatter}

\bibliography{biblio}

\end{document}